\documentclass[11pt]{article}
\usepackage{epsfig,amssymb,graphicx,amsmath,amsthm,subcaption}
\usepackage{authblk}
\usepackage{array}

\newcommand{\ave}[1]{\ensuremath{\langle {#1} \rangle}}
\newcommand{\pd}[2]{\ensuremath{\frac{\partial {#1}}{\partial {#2}}}}
\newcommand{\diff}[2]{\ensuremath{\frac{d {#1}}{d {#2} }}}

\begin{document}

\title{Mean-field models for non-Markovian epidemics on networks: from edge-based compartmental to pairwise models}
\author{N. Sherborne$^{\rm 1}$, J.C. Miller$^{\rm 2}$,
	K.B. Blyuss$^{\rm 1}$\footnote{Corresponding author: K.Blyuss@sussex.ac.uk}, I.Z. Kiss$^{\rm 1}$
}
\affil{$^{\rm 1}$ Department of Mathematics, School of Mathematical and Physical Sciences,
University of Sussex, Falmer, Brighton BN1 9QH, UK}

\affil{$^{\rm 2}$ School of Mathematics, School of Biology, and MAXIMA, Monash University, Melbourne, VIC Australia
and Institute for Disease Modeling, Bellevue, Washington 98005, USA}

\maketitle

\begin{abstract}
This paper presents a novel extension of the edge-based compartmental model for epidemics with arbitrary distributions of transmission and recovery times. Using the message passing approach we also derive a new pairwise-like model for epidemics with Markovian transmission and an arbitrary recovery period. The new pairwise-like model allows one to formally prove that the message passing and edge-based compartmental models are equivalent in the case of Markovian transmission and arbitrary recovery processes. The edge-based and message passing models are conjectured to also be equivalent for arbitrary transmission processes; we show the first step of a full proof of this. The new pairwise-like model encompasses many existing well-known models that can be obtained by appropriate reductions. It is also amenable to a relatively straightforward numerical implementation. We test the theoretical results by comparing the numerical solutions of the various pairwise-like models to results based on explicit stochastic network simulations.
\end{abstract}

\section{Introduction}
The use of mathematical tools to study and understand the spread of infectious diseases is an established and fruitful area of research. In their 1927 paper \cite{kermack1927contribution} Kermack and McKendrick established the susceptible-infected-recovered (SIR) framework which forms the basis of many models to this day. However, these early models do not consider realistic human behaviour and interactions. A particular challenge for the construction of more realistic models lies in capturing these contact patterns. Typically, in large populations each individual only interacts with a few others, and these connections determine the possible routes of disease transmission.

Moreover, studies have found significant heterogeneity in the number of contacts a single individual may have \cite{pastor2001epidemic}, as well as a tendency for two connected nodes to share further common neighbours. This clustering of individuals introduces multiple potential routes of transmission between any two individuals and provides a challenge to modellers \cite{watts1998collective,volz2011effects}.

A major development which takes into account these factors has been the use of graphs or networks which provide a detailed description of these contact patterns, as opposed to assuming that each individual interacts directly with every other member of the population. In network-based models individuals are represented by nodes in the network, with edges (or links) encoding the interactions between the nodes.

The resulting models of epidemics on such networks lead to a continuous time Markov chain on a very large but discrete state space. The direct analysis of such stochastic models for realistic networks is out of reach, unless strong simplifying assumptions about the networks are made. However, alternatives to deal with such models do exist. For example, it is possible to derive appropriate mean-field models that are capable of capturing the average behaviour of a stochastic model with a high degree of accuracy. These then allow one to analytically determine, explicitly or implicitly, quantities, such as the epidemic threshold or final epidemic size, and reveal how these depend on the characteristics of the network.

Deriving such deterministic mean-field models can be done in several different ways, depending on how the averaging is done. For example, averaging over all nodes and links in all possible states leads to pairwise models \cite{keeling1999,house2011insights}, while considering each individual separately and all possible ways in which it can become infected by its neighbours leads to the message passing formalism \cite{karrer2010message}. Furthermore, choosing to average over all possible star-like structures that are typically defined by a node and all its neighbours, and also taking into account their disease status, yields the so-called effective-degree models \cite{lindquist2011effective}. One of the most compact mean-field models is the edge-based compartmental model which is based on considering a randomly chosen test node and working out its probability of staying susceptible, this probability is equivalent to the proportion of nodes that are susceptible in the entire population \cite{miller2012edge}, see \cite{danon2011networks,pastor2015epidemic,kiss2016mathematics} for reviews. Of course, all of these models start from the same exact stochastic model but choose to average over a different scale, thus it is not surprising that some of these models are, in fact, equivalent \cite{house2011insights,taylor2014interdependency,miller2014epidemic,kiss2016mathematics}, as we will demonstrate later on.

Throughout this paper we restrict ourselves to unweighted, bi-directional and static networks constructed according to the configuration model (CM) \cite{bender1978asymptotic}. Every node $u$ is assigned a number of neighbours, known as its degree, according to a probability distribution $p_k$ known as the {\it degree distribution}. This describes the probability of a randomly chosen node having degree $k$. One can also define the following generating functions which will be used throughout the paper
\begin{equation}
\begin{split}
	& G_0(x) := \sum_k p_k x^k, \quad G_1(x) := \frac{1}{\ave{k}} \sum_k p_k k x^{k-1}, \\
	& G_2(x) := G_1'(x) = \frac{1}{\ave{k}} \sum_k p_k k(k-1) x^{k-2},
\label{eq:genFunc}
\end{split}
\end{equation}
where $\ave{k}=G_{0}^{'}(1)$ is the mean degree. $G_1(x)$ is the generating function for the {\it excess degree distribution}, since $kp_k/\langle k \rangle$ describes the probability that a node reached by traversing a randomly selected edge has $(k-1)$ other contacts \cite{newman2002spread}.  On a tree network with no short loops the number of neighbours that an infected node can reach is determined by the excess degree of the nodes; the mean excess degree is given by $G_2'(1)$. The moment generating function $G_2(x)$ is used to trace the route of infection.

In addition to describing the underlying contact pattern, we must consider the disease characteristics. These are the force of infection or transmission and the typical length of time between an individual becoming infected and later recovering to (permanent) immunity (henceforth referred to as the recovery time). Many models assume that the underlying stochastic transmission and recovery processes are memoryless \cite{keeling2005networks,volz2008sir,house2011insights}. These assumptions lead to models which are mathematically tractable and relatively simpler to analyse when compared to models where the time to infection or the recovery time are chosen from arbitrary distributions other than the exponential.

However, when compared to data, these assumptions are often violated, and diseases can exhibit unique and non-Markovian behaviour in terms of the strength and duration of infection. For example, recovery time distributions are usually better approximated by some peaked distribution with a well defined mean, see e.g. \cite{bailey1954statistical,gough1977estimation,wearing2005appropriate} and references therein. Some work has been done recently on modelling network epidemics with non-exponential distribution of infectious periods \cite{kiss2015pairwise,sherb15,sherb16}. In this paper we will take the most general view possible by letting $\tau(a)$ and $q(a)$ denote the general probability density functions of the time to infection across a link between an infectious and susceptible node, and the recovery time of an infected/infectious node, respectively. Here, $a$ denotes the time since the node became infected, also known as the age of infection. Once a susceptible node has been exposed to a transmission event it becomes infected  immediately - a latent period can be included by setting the transmission distribution $\tau(a)$ to be zero for some initial period of time. Recovery from the disease grants lifetime immunity. Using these distributions assumes a homogeneous response to disease; whilst this restriction is not always necessary (see e.g. \cite{wilkinson2014message}), it is a common simplification in order to obtain a concise model. 

Considering locally tree-like networks with arbitrary heterogeneous degree distributions, and a stochastic epidemic model where both the transmission and recovery processes are as general as possible, we set out to derive the most general edge-based and pairwise-like models, and to establish different model equivalences and relationships between these and existing models.

The rest of the paper is organised as follows: in the following section we describe how the message passing (MP) approach works \cite{karrer2010message} and how the resulting model is constructed. We also present a novel extension of the edge-based compartmental model (EBCM) \cite{miller2012edge} to networks with general degree distributions and for SIR epidemics with arbitrary transmission and recovery processes. In Section~\ref{sec:derive}, starting from the message passing model and considering Markovian transmission, we show how one can derive a low-dimensional or compact pairwise-like model (PLM) which is more amenable to implement numerically but is still capable of modelling arbitrary recovery time and degree distributions. In Section~\ref{sec:equivalence} we use the new PLM to show equivalence between the MP model and EBCM for Markovian transmission, and thus to conjecture that the MP model and EBCM are equivalent in a general setting. The PLM is also reducible to many existing models for specific choices of the recovery time or degree distribution, as will be shown in Section~\ref{sec:reduction}. In Section~\ref{sec:sims} we compare the numerical solutions of the mean-field models to averaged results from explicit stochastic network simulations. The good agreement between these confirm and support earlier theoretical findings. The paper concludes with a brief discussion of the major results and possible future work.

\section{Model summary}
\label{sec:models}
\subsection{The message passing (MP) method}
\label{sec:message}

In their 2010 paper \cite{karrer2010message} Karrer and Newman introduced the message passing approach to model SIR dynamics on networks. Here, we briefly present the ideas behind their model and its assumptions. Recalling $\tau(a)$ and $q(a)$ as the densities for transmission and recovery time one can introduce a new function $f(a)$
\begin{equation}
	f(a) = \tau(a) \int_{a}^{\infty} q(x) dx,
	\label{eq:fDist}
\end{equation}
such that the probability that an infected node attempts to transmit the disease to a given neighbour before time $t$ is $\int_0^t f(a) da$, since a neighbour can only transmit the disease if it has not yet recovered. Note that the result of integration of (\ref{eq:fDist}) over all time is equal to the overall probability of the disease being transmitted across a given network edge, commonly known as the {\it transmissibility}. This is an important quantity which is used in percolation models to determine the epidemic threshold and expected final epidemic size for a major outbreak \cite{newman2002spread,kenah2007second}. 

In order to model the dynamics of disease spread consider a test node $u$. This node is placed into a cavity state where it can become infected but is not able to transmit the disease to any of its neighbours. This has no effect on the probability of the node being in any given state \cite{miller2012edge}. Now consider a node $v$ which is a neighbour of $u$; the {\it message} is the probability that $v$ has {\it not} transmitted the disease to node $u$ by time $t$, denoted $H^{u \leftarrow v}(t)$. This probability is comprised of two distinct possibilities; the first possibility is that $v$ makes no attempt to transmit the disease before $t$ regardless of its own age of infection, given by $1 - \int_0^t f(a) da$. Alternatively, it could be that $v$ will transmit to $u$ at age $a <t$, but $v$ itself was infected at some time $t_1 > t - a$ and has, therefore, not yet attempted to transmit the disease to its neighbour $u$. This requires $v$ to have initially been susceptible (with probability $z$) and to have escaped transmission from each of its neighbours (excluding $u$) until at least time $(t-a)$, which is exactly $z \int_0^t f(a)\prod_{w \in \mathcal{N}(v)\backslash u} H^{v \leftarrow w}(t-a) \,da$, where $\mathcal{N}(v)$ denotes the set of neighbours of $v$. Hence, combining these two gives
\begin{equation}
	H^{u \leftarrow v}(t) = 1 - \int_0^t f(a) \left[ 1 - z \prod_{w \in \mathcal{N}(v)\backslash u} H^{v \leftarrow w}(t-a) \right] da.
	\label{eq:Huv}
\end{equation}
In principle, one could calculate (\ref{eq:Huv}) for all edges (in both directions) to find a full solution for the proportion of the population that is susceptible, infected or removed at any time $t$. For example, the probability $u$ is susceptible is the product of $H^{u \leftarrow w}(t)$ across all neighbours $w \in \mathcal{N}(u)$ multiplied by the probability that it was initially susceptible, $z$. On a single fixed finite tree network, solving (\ref{eq:Huv}) for all edges will, in fact, yield the exact solution of the stochastic epidemic \cite{karrer2010message}. The size of such a system of equations would be twice the number of all edges in the network (since both $H^{u \leftarrow v}(t)$ and $H^{v \leftarrow u}(t)$ would need to be calculated).

However, for CM networks it is possible to modify (\ref{eq:Huv}) so that it performs as an averaged probability of all edges across the whole network, this is denoted $H_1(t)$ \cite{karrer2010message}. In order to account for degree heterogeneity, the product in (\ref{eq:Huv}) is replaced by the excess degree distribution $G_1(H_1(t-a))$. Thus, we now have
\begin{equation}
	H_1(t) = 1 - \int_0^t f(a) \left[ 1 - z G_1(H_1(t - a)) \right] \, da,
	\label{eq:H1}
\end{equation}
with $H_1(0) = 1$. Now the somewhat difficult part is to obtain  and  solve, at least numerically, the differential or integro-differential equation for $H_1(t)$. For the purely Markovian case, with transmission and recovery parameters $\beta$ and $\gamma$ respectively, this is
\begin{equation*}
	\frac{dH_1}{dt} = \gamma - (\beta + \gamma)H_1(t) + \beta z G_1(H_1(t)),
\end{equation*}
where $z$ is the fraction of the population which was initially susceptible at time $t=0$ \cite{karrer2010message}. However, the precise form of this equation is not universal, it depends on the particular choice of the transmission and recovery processes. The proportion of susceptible, infected and recovered individuals at any time $t$ are then given, in terms of the message $H_1(t)$, as
\begin{equation}\label{eq:msg_system}
\begin{split}
	\ave{S}(t) &= z G_0(H_1(t)), \\
	\ave{R}(t) &= \int_0^t q(a)\left[1 - \ave{S}(t - a)\right] \, da, \\
	\ave{I}(t) &= 1 - \ave{S}(t) - \ave{R}(t).
\end{split}
\end{equation}

For CM networks, as the size of the network increases to infinity, the length of the shortest loops also diverges to infinity, and, therefore, the network becomes locally tree-like. The result of this is that an MP model with the average message $H_1$ is exact when the stochastic epidemic is considered on the ensemble of CM networks \cite{karrer2010message}. Although this approach is theoretically able to model dynamics for arbitrary choices of transmission and recovery processes, the need to find a numerically solvable differential equation for $H_1$ in (\ref{eq:H1}) has restricted the use of MP, and numerical examples are limited, see \cite{karrer2010message,wilkinson2014message} for several examples where output from the MP model is compared to results based on simulations.

\subsection{EBCM for general transmission and recovery processes}

The edge-based compartmental model has been used for Markovian dynamics \cite{miller2012edge}. We introduce a new extended EBCM which generalises the method to general transmission and recovery processes. 

The EBCM uses the instantaneous rates of transmission and recovery given by the {\it hazard functions} rather than the raw densities $\tau(a)$ and $q(a)$. These are defined as
\begin{equation}
	\zeta(a) := \frac{\tau(a)}{\xi_\tau(a)}, \quad \text{and} \quad \rho(a) := \frac{q(a)}{\xi_q(a)},
	\label{eq:hazards}
\end{equation}
where $\xi_\tau(a)$ and $\xi_q(a)$ are the respective {\it survival functions}
\begin{equation}
	\xi_\tau(a) = \int_a^{\infty} \tau(\hat{a}) \, d\hat{a}, \quad \text{and} \quad \xi_q(a) = \int_a^{\infty} q(\hat{a}) \, d\hat{a}.
	\label{eq:survival}
\end{equation}

It is possible to express these survival functions in an alternative form which will be useful later. These are given in the following lemma.\\

\noindent {\bf Lemma 1.} {\it Given $\xi_{\tau}(a)$ and $\xi_q(a)$ as defined in (\ref{eq:survival}), and the relations in (\ref{eq:hazards}), one can write}
	\begin{equation}
		\xi_{\tau} (a) = e^{\displaystyle{-\int_0^a \zeta(\hat{a})\, d\hat{a}}} \quad \text{and} \quad \xi_{q} (a) = e^{\displaystyle -\int_0^a \rho(\hat{a}) \, d\hat{a}}.
	\end{equation}

\noindent {\bf Proof.}
	From (\ref{eq:hazards}) and (\ref{eq:survival}) we see that
	\[ \xi_{\tau}(a) = \int_a^{\infty} \zeta(\hat{a})\xi_{\tau}(\hat{a}) \, d\hat{a}.\]
	Differentiating this equation with respect to $a$ yields a first order differential equation
	\[ \diff{\xi_{\tau}}{a} = - \zeta(a)\xi_{\tau}(a), \] 
	whose solution with the initial condition $\xi_\tau(0)=1$ gives
	\[
		\xi_{\tau}(a) = e^{\displaystyle -\int_0^{a} \zeta(\hat{a}) \, d\hat{a}}.
	\]
	Applying the same steps to $\xi_q(a)$ completes the proof. \hfill$\blacksquare$\\

All of these disease variables and related functions are summarised in Table~\ref{tab:diseaseVar}. As before, the contact network is a CM network with degree distribution and generating functions as defined in (\ref{eq:genFunc}). The basis of the EBCM revolves around finding the probability that a random test node is in a susceptible, infected or recovered state at time $t$; as this test node is chosen at random these probabilities are equal to the proportions of the population in each state at time $t$, denoted $S(t)$, $I(t)$ and $R(t)$ respectively. Again, the test node is placed into a cavity state, and the probability of remaining susceptible until time $t$ is the probability of the test node being initially susceptible and escaping transmission from each of its neighbours up to time $t$. This concept is similar to the notion and use of $H_1$ in MP models. Recovery is modelled using age-structured differential equations.

{
	\setlength{\extrarowheight}{6pt}
	
	\begin{table}
		\begin{center}
			\begin{tabular}{|m{0.11\textwidth}|m{0.8\textwidth}|}
				\hline
				{\textbf{Variable}} & {\textbf{Definition}}\\\hline
				{$\tau(a)$} & {The density of the transmission process.}\\ 
				\hline
				{$q(a)$} & {The density of the recovery process.}\\
				\hline
				{$\xi_{\tau}(a)$} & {The \emph{survival function} of the transmission process. The probability that an infected node of age $a$ has not yet attempted to transmit the disease along a given edge: $\int_a^{\infty} \tau(x) \, dx$.}\\
				\hline
				{$\xi_{q}(a)$} & {The \emph{survival function} of the recovery process. The probability that an infected node reaches at least age $a$ before recovering: $\int_a^{\infty} q(x) \, dx$.}\\ 
				\hline
				{$\zeta(a)$} & {The \emph{hazard function} of the transmission process. The probability of an edge of age $a$ transmitting in a small interval of time $(a, a+\Delta a)$: $\frac{\tau(a)}{\xi_{\tau}(a)}$.}\\ 
				\hline
				{$\rho(a)$} & {The \emph{hazard function} of the recovery process. The probability of an infected node of age $a$ recovering in a small interval of time $(a, a+\Delta a)$: $\frac{q(a)}{\xi_{q}(a)}$.}\\ 
				\hline
				{$f(a)$} & {The probability that, in a small interval, an infectious contact is made by an infected node of age $a$ :$\tau(a)\int_a^{\infty} q(x) \, dx$.}\\ 
				\hline
				{$g(a)$} & {The probability that, in a small interval, an infectious node of age $a$ recovers, without attempting to transmit the disease to a given neighbour: $q(a)\int_a^{\infty} \tau(x) \, dx$.}\\ 
				\hline				
			\end{tabular}
		\end{center}
		\caption{The variables and functions describing the transmission and recovery processes.}
		\label{tab:diseaseVar}
	\end{table}
}

{
	\setlength{\extrarowheight}{4pt}
	
	\begin{table}
		\begin{center}
			\begin{tabular}{|m{0.11\textwidth}|m{0.8\textwidth}|}
				\hline
				{\textbf{Variable}} & {\textbf{Definition}}\\\hline
				{$\Theta(t)$} & {The probability an initially susceptible test node has not received a transmission from a random neighbour by time $t$.}\\
				\hline
				{$\Phi_S(t)$} & {The probability that a random neighbour of a test node $u$ is still susceptible. }\\
				\hline
				{$\Phi_I(t)$} & {The probability that a random neighbour of a test node $u$ is infected, but has not transmitted to $u$.}\\
				\hline
				{$\phi_I(t,a)$} & {The density for a random neighbour of a
					test node $u$ to be infected, have not transmitted to $u$ by time $t$
					and be $a$ units of time into its infection,  $\Phi_I(t) = \int_0^t
					\phi_I(t,a) \, \mathrm{d}a$.}\\
				\hline
				{$\Phi_R(t)$} & {The probability a random neighbour of a test node $u$ has been infected and recovered without transmitting to $u$.}\\
				\hline
				{$S(t)$} & {The
					density of susceptible nodes.}\\
				\hline
				{$I(t)$} & {The
					density of infected nodes.}\\\hline
				{$i(t,a)$} & {The
					density of infected nodes that were infected at time
					($t-a$).}\\\hline
				{$R(t)$} & {The
					density of recovered nodes.}\\ \hline
				{$G_1(x)$} & {The
					probability generating function of the excess degree distribution:
					$\frac{1}{\ave{k}}\sum_{k=0}^\infty p_k k x^{(k-1)}$.}\\\hline
				{$G_2(x)$} & {The derivative of the
					probability generating function of the excess degree distribution:
					$\frac{1}{\ave{k}}\sum_{k=0}^\infty p_k k(k-1) x^{(k-2)}$.}\\\hline
			\end{tabular}
		\end{center}
		\caption{The list of variables in the EBCM.}
		\label{tab:nonMarkovEBCMvars}
	\end{table}
}

The first important quantity is $\Theta(t)$, defined in a manner similar to $H_1(t)$ in (\ref{eq:H1}) as the probability that the representative test node has not received transmission from a given neighbour by time $t$. This approach then differs from MP by directly expressing a differential equation for the dynamics of $\Theta$. The model is known as ``edge-based" because it considers the state of the neighbours of the test node; the densities $\Phi_S(t)$, $\Phi_I(t)$ and $\Phi_R(t)$ describe the probability that at time $t$ a random neighbour of the test node is (i) still susceptible, (ii) infected but has not attempted to transmit the disease to the test node, (iii) recovered, and it did not transmit to the test node whilst it was infected. Since the age of infection is crucial in determining the hazard rates we also introduce the variables $i(t,a)$ and $\phi_i(t,a)$ as the densities of infected nodes with the age of infection $a$, and infected neighbours who have not transmitted to the test node and have age $a$, respectively. Thus, it is clear that $I(t) = \int_0^t i(t,a) da$ and $\Phi_I(t) = \int_0^t \phi_I (t,a) da$. These variables are summarised in Table~\ref{tab:nonMarkovEBCMvars}. The new EBCM is presented below for the case where the disease is introduced by randomly infecting a chosen fraction $1-z$ of the population at time $0$.
\begin{subequations}
\begin{align}
	\Phi_S(t) &= zG_1(\Theta(t)),\\
	\phi_I(t,0) &= - \dot{\Phi}_S(t), \nonumber \\
				&= (1-z)\delta(t) + zG_2(\Theta(t)) \int_0^t \zeta(a) \phi_I(t,a) \, d a, \label{eq:EBCM:phiI}\\
	\left(\pd{}{t} + \pd{}{a} \right) \phi_I(t,a) &= - \left[\zeta(a) + \rho(a)\right]\phi_I(t,a), \qquad 0 < a \leq t, \label{eq:EBCM:phiDiff}\\
	\Phi_I(t) &= \int_0^t \phi_I(t,a) \, d a, \label{eq:EBCM:PhiInt}\\
	\diff{\Theta(t)}{t} &= -\int_0^t \zeta(a) \phi_I(t,a) \, d a, \label{eq:EBCM:Theta}\\
	\Phi_R(t) &= \Theta - \Phi_S - \Phi_I,\\
	S(t) &= zG_0(\Theta(t)), \label{eq:EBCM:S}\\
	\left(\pd{}{t} + \pd{}{a} \right) i(t,a) &= - \rho(a) i(t,a), \qquad 0 < a \leq t, \label{eq:EBCM:iAge} \\
	i(t,0) &= - \dot{S}(t), \nonumber \\
		& = (1-z)\delta(t) + \ave{k}zG_1(\Theta(t))\int_0^t \zeta(a) \phi_I(t,a) \, d a, \label{eq:EBCM:iBdry}\\
	I(t) &= \int_0^t i(t,a) \, da, \label{eq:EBCM:I}\\
	R(t) &= 1-S-I.
\end{align}
	\label{eq:nonMarkEBCM}
\end{subequations}
Where $\delta(t)$ is the Dirac $\delta$-function, and we have the initial condition
\begin{equation*}
	\Theta(0) =1,
\end{equation*}
and where $\ave{k}$ denotes the average node degree.

Compared to the MP method, the new EBCM is more universal in the sense that different choices of $\tau(a)$ and $q(a)$ can be substituted straight into the general model rather than having to consider how the equation for $H_1$ changes as a result of the choices being made. This arguably makes the resultant EBCM more amenable to numerical implementation, especially in cases where the evolution equation for $H_1(t)$ is difficult to obtain. 

As it turns out, setting up the model in such a general way allows one to derive an implicit analytical relation for the final epidemic size, as presented in the following result.

\noindent {\bf Theorem 1.} {\it The final size of the epidemic $r_\infty = R(\infty)$ in the EBCM~(\ref{eq:nonMarkEBCM}) with a vanishingly small proportion of
infected nodes at time $t=0$ is
\[
	r_\infty = 1- G_0(\Theta_\infty),
\]
where $\Theta_\infty$ solves the equation
\[
	\Theta_{\infty} = 1-\widetilde{T} + \widetilde{T} G_1(\Theta_{\infty}),
\]
and $\widetilde{T} =\int_0^\infty \zeta(a) \exp(-\int_0^a
\left[\zeta(\hat{a}) + \rho(\hat{a}) \right] \, d\hat{a})\, da$ is known as the {\it transmissibility} of the disease. It is the probability that the disease is transmitted along an edge (in isolation) before recovery.}\\

\noindent {\bf Proof.} From (\ref{eq:EBCM:S}),  for $z \to 1$ and $t \to \infty$ it immediately follows that
\[ 
	r_{\infty} = 1 - S(\infty) = 1 - G_0(\Theta(\infty)).
\]
Furthermore, we have
\[
	\Theta(\infty) = 1- \int_0^\infty \int_0^t \zeta(a) \phi_I(t,a) \,
	da \, dt.
\]
Interchanging the order of integration yields
\[
	\Theta(\infty) = 1 - \int_0^\infty \int_a^\infty \zeta(a) \phi_I(t,a)
\, dt \, da.
\]
Setting $u=t-a$ and noting that $\phi_I(t,a) = \phi_I(u,0)\exp\left(-\int_0^a
	\left[\zeta(\hat{a})+\rho(\hat{a})\right]\, d\hat{a}\right)$ yields
\begin{align*}
	\Theta(\infty) &= 1 - \int_0^\infty \int_0^\infty \phi_I(u,0) \zeta(a)
e^{-\int_0^a \left[\zeta(\hat{a})+\rho(\hat{a})\right] \,d\hat{a}} \,
du \, da\\
	&= 1- \left[\int_0^\infty \phi_I(u,0) du\right] \int_0^\infty \zeta(a)
e^{-\int_0^a \left[\zeta(\hat{a})+\rho(\hat{a})\right] \, d\hat{a}}\, da\\
	&= 1+ \left[\int_0^\infty \dot{\Phi}_S(u) \, du\right] \int_0^\infty \zeta(a)
e^{-\int_0^a \left[\zeta(\hat{a})+\rho(\hat{a})\right] \, d\hat{a}} \, da\\
	&= 1 + (\Phi_S(\infty)-\Phi_S(0) ) \widetilde{T}\\
	&= 1 + \widetilde{T} G_1(\Theta(\infty)) -\widetilde{T}. 
\end{align*}\hfill$\blacksquare$

The result in Theorem 1 corresponds to well-known results based on tools from percolation theory \cite{newman2002spread,kenah2007second,miller2007epidemic}, and it is equivalent to the final epidemic size obtained for MP models \cite{karrer2010message}. This provides good evidence that the new EBCM (\ref{eq:nonMarkEBCM}) is an accurate representation of the true expected dynamics of an SIR epidemic starting from a vanishingly small number of initially infected seeds.

\section{Derivation of the new pairwise-like model}
\label{sec:derive}

The number of individuals becoming infected is related to the number of edges connecting susceptible nodes to infected neighbours. Pairwise models traditionally construct differential equations for the expected numbers of such edges, which themselves depend on the numbers of triples in certain states (e.g. susceptible-susceptible-infected). To break this dependence a moment closure approximation is commonly used to express the number of triples in terms of pairs and individuals \cite{keeling1999}. Recently, Wilkinson and Sharkey \cite{wilkinson2014message} and Wilkinson et al. \cite{wilkinson2016relationships} have shown that for regular tree networks exact pairwise models can be derived from the MP model when the transmission process is assumed to be Markovian. Here we use similar methods with the notation from Section~\ref{sec:message} to extend this result to heterogeneous networks.

Firstly, we define the new variable $\ave{SI}(t)$ as the proportion of edges in the network which connect a susceptible node to an infected one at time $t$. This can be defined in terms of existing quantities; first the susceptible node must have been initially susceptible and escaped infection from all other neighbours, given by $zG_1(H_1(t))$. This must be multiplied by the probability that the remaining neighbour of the susceptible node is infected and has not yet transmitted the disease to this neighbour. To find this probability it is easier to calculate all other possibilities and subtract them from one. These possibilities are: (i) the neighbour is still susceptible, (ii) the neighbour has already transmitted the disease, (iii) the neighbour was infected but recovered without transmitting the disease. Combining these gives
\begin{subequations}
\begin{align}
	\ave{SI}(t) &= z G_1(H_1(t))\left[\vphantom{\int_0^t} 1 - zG_1(H_1(t))\right. \nonumber \\
	&\phantom{z G_1(H_1(t)) (1 -)}\left. - \int_0^t \{f(a) + g(a) \}\left[1 - z G_1(H_1(t - a))\right] \,da\right], \label{eq:SIrelation1}\\
	&= z G_1(H_1(t))\left[\vphantom{\int_0^t} H_1(t) - zG_1(H_1(t)) \right. \nonumber \\
	&\phantom{z G_1(H_1(t)) (1 -)} - \left.\int_0^t g(a)\left[1 - zG_1(H_1(t-a))\right] \, da\right], \label{eq:SIrelation2}\\
\end{align}
\label{eq:pairs}
\end{subequations}
where
\begin{equation}
	g(a) := q(a)\int_{a}^{\infty} \tau(x) dx,
	\label{eq:gt}
\end{equation}
is the probability of an infected node recovering in the interval $(a, a + \Delta a)$ without transmitting to a given neighbour. The corresponding population-level quantity is given by
\begin{equation}
	[SI](t) = \ave{k} N \ave{SI}(t),
	\label{eq:SInum}
\end{equation}
where $N$ denotes the total size of the population.

We now assume that the time to transmission is exponentially distributed, i.e. $\tau(a) = \beta e^{-\beta a}$. Substituting this into (\ref{eq:H1}), changing the variable to $t' = t - a$, and using the Leibniz rule gives
\begin{align}
	\diff{H_1}{t} &= -\beta\left[ 1 - z G_1(H_1(t)) - \int_0^t q(t - t')e^{-\beta (t-t')}\left[1 - zG_1(H_1(t'))\right] \, da \right. \nonumber \\
	&\hspace{1.5cm} - \left.\int_0^{t} \beta e^{-\beta (t - t')} \left(\int_{t - t'}^{\infty} q(x) dx\right)\left[1 - zG_1(H_1(t'))\right] \, dt' \right] \nonumber \\
	&= -\beta\left[ 1 - zG_1(H_1(t)) - \int_0^t \{f(a) + g(a) \}\left[1 - zG_1(H_1(t'))\right] \, da\right] \nonumber\\
	&= -\beta\frac{\langle SI \rangle(t)}{z G_1(H_1(t))} \\
	&= -\beta\frac{[SI]}{z \ave{k} N G_1(H_1)}. \label{eq:H_1dot}
\end{align}

For the infected population, using (\ref{eq:msg_system}) and identities, such as $[S](t) = N\ave{S}(t)$, leads to
\begin{align}
	\dot{[I]} &= - \dot{[S]} - \dot{[R]} \\
	&= \beta [SI] - \beta \int_0^t q(a) [SI](t-a) da - q(t)N(1-z). \label{eq:PLMIdot}
\end{align}
The majority of pairwise epidemic models retain an explicit differential equation for the prevalence \cite{house2011insights,wilkinson2016relationships}. However, we choose to integrate (\ref{eq:PLMIdot}) to reduce the number of differential equations which must be integrated numerically. By noting that $[SI] = \beta\int_0^t \dot{[SI]}(t-a)\xi_q(a) \, da$ and $q(a) = -\xi_q'(a)$ we have
\begin{equation*}
	\dot{[I]} = \beta \int_0^t \left( \dot{[SI]}(t-a)\xi_q(a) + \xi_q'(a) [SI](t-a) \right)\, da - q(t)N(1-z),
\end{equation*}
which is the result of differentiating
\begin{equation}
	[I] = \beta\int_0^t [SI](t-a)\xi_q(a) \, da + N(1-z)\xi_q(t).
	\label{eq:PLMI}
\end{equation}
Whilst (\ref{eq:PLMIdot}) facilitates easier comparison to existing models we will use its equivalent representation (\ref{eq:PLMI}) for computational efficiency.

For the variable $\langle SI \rangle$ the calculation is more laborious; working term-by-term from~(\ref{eq:SIrelation2}) and using the new relation (\ref{eq:H_1dot}) one obtains
\begin{align*}
	\dot{\langle SI \rangle} &= -z G_2(H_1)\left(\beta\frac{[SI]}{z \ave{k} N G_1(H_1(t))}\right)\Bigg[\cdot \Bigg] \\
	&\phantom{= z} - \beta\frac{[SI]}{\ave{k}N} + z \beta [SI] \frac{G_2(H_1(t))}{\ave{k}N} - q(t)e^{-\beta t}(1-z)z G_1(H_1(t)) \\
	&\phantom{= z} -  z \beta \int_0^t q(a)e^{-\beta a}[SI](t-a)G_2(H_1(t-a)) 	\frac{G_1(H_1(t))}{N\ave{k}G_1(H_1(t-a))} \, da,
\end{align*}
where $[\cdot]$ denotes the large bracket in~(\ref{eq:SIrelation2}), and the Leibniz rule has been used again to resolve the integral term. Finally, based on~(\ref{eq:pairs}), $[\cdot]=\frac{\langle SI \rangle}{zG_{1}(H_1)}$, which allows us to eliminate $[\cdot]$ and replace it with a term involving $\langle SI \rangle$. Then, multiplying through by $\ave{k}N$ one obtains $\dot{[SI]}$ in (\ref{eq:newSystem}) below.
\begin{samepage}
\begin{subequations}
\begin{align}
	\dot{H_1} &= -\beta \frac{[SI]}{z \ave{k} N G_1(H_1)}, \\
	\dot{[SI]} &= -\beta [SI][SI]\frac{G_2(H_1)}{z\ave{k}N [G_1(H_1)]^2}  - \beta[SI] \nonumber \\
	&\phantom{-\beta} + z \beta [SI]G_2(H_1) -  q(t)e^{-\beta t}(1-z)z G_1(H_1) \ave{k} N \label{eq:PLM:SI}\\
	&\phantom{-\beta} -  z \beta \int_0^t q(a)e^{-\beta a} [SI](t-a)G_2(H_1(t-a)) \frac{G_1(H_1(t))}{G_1(H_1(t-a))} \, da , \nonumber \\
	[I] & = \beta\int_0^t [SI](t-a)\xi_q(a) \, da + N(1-z)\xi_q(t).
\end{align}
	\label{eq:newSystem}
\end{subequations}
\end{samepage}

At any time $t$ the expected number of susceptibles can be found as $[S](t) = zN G_0(H_1(t))$. We call the system (\ref{eq:newSystem}) the pairwise-like model (PLM). The major benefits of this model are its compact size, requiring only two differential equations to be solved, and the fact that retaining the concept of the ``message" from the MP model has meant that moment closure approximations are not necessary. The PLM is a concise, flexible and exact model for SIR dynamics on CM networks. Moreover, the numerical solution of such an ODE system is straightforward to compute.

There are elements of (\ref{eq:newSystem}) that are similar to recent models \cite{house2011insights,wilkinson2016relationships,rost2016pairwise}, which helps us identify links between different methodologies and promote a greater and more unified understanding of this area. These connections are explored in detail in Section~\ref{sec:reduction}. 

\section{Model equivalence}
\label{sec:equivalence}

Before exploring how the new PLM (\ref{eq:newSystem}) reduces to recent and classical models, we use it to show equivalence between the MP model and the EBCM introduced in Section~\ref{sec:models} in the special case of Markovian transmission but an arbitrary recovery process. Since the PLM is derived directly from the MP model, it is only necessary to show equivalence between the PLM and EBCM. Note that whilst all variables in both the MP model and EBCM are based on proportions, the PLM deals with population-level quantities, and hence we will show that the corresponding quantities (e.g. proportion and population of infection) share identical dynamics and differ only by constants, such as the total size of the population, $N$. This equivalence is summarised in the following theorem.\\

\noindent {\bf Theorem 2.} {\it Given an exponentially distributed transmission process such that $\tau(a) = \beta e^{-\beta a}$ with some $\beta >0$, the dynamics of the EBCM (\ref{eq:nonMarkEBCM}) are equivalent to that of the PLM and, therefore, are identical to the MP model and stochastically exact on CM networks as the population size tends to infinity.}\\

\noindent {\bf Proof.} The PLM (\ref{eq:newSystem}) has been derived directly from the MP model (\ref{eq:msg_system}), so to prove equivalence between the MP and EBCM systems it will suffice to show that the PLM can be independently derived from the EBCM (\ref{eq:nonMarkEBCM}). We introduce the new variables
\begin{subequations}
	\begin{align}
	\{ SI \} &= z N \ave{k} G_1(\Theta(t)) \Phi_I(t), \label{eqn:curlySIdef}\\
	\{I\} &= NI, \label{eqn:curlyIdef}
	\end{align}
	We then derive equations for $\{\dot{SI}\}$ and $\{\dot{I}\}$ and show that 
	\begin{equation}
	\diff{}{t}\Theta = -\beta\frac{\{SI\}}{z\ave{k}NG_1(\Theta(t))}. \label{eqn:curlyThetadiff}
	\end{equation}
\end{subequations} 
The resulting system of equations will be identical to those of system~\eqref{eq:newSystem} with $\Theta$ playing the role of $H_1$, $\{SI\}$ playing the role of $[SI]$ and $\{I\}$ playing the role of $[I]$. Thus the PLM and EBCM systems are equivalent.

The relation~\eqref{eqn:curlyThetadiff} immediately follows from $\diff{\Theta}{t} = -\beta \Phi_I$ and substituting for $\Phi_I$ in terms of $\{SI\}$.

\begin{figure}
\includegraphics[width = \textwidth]{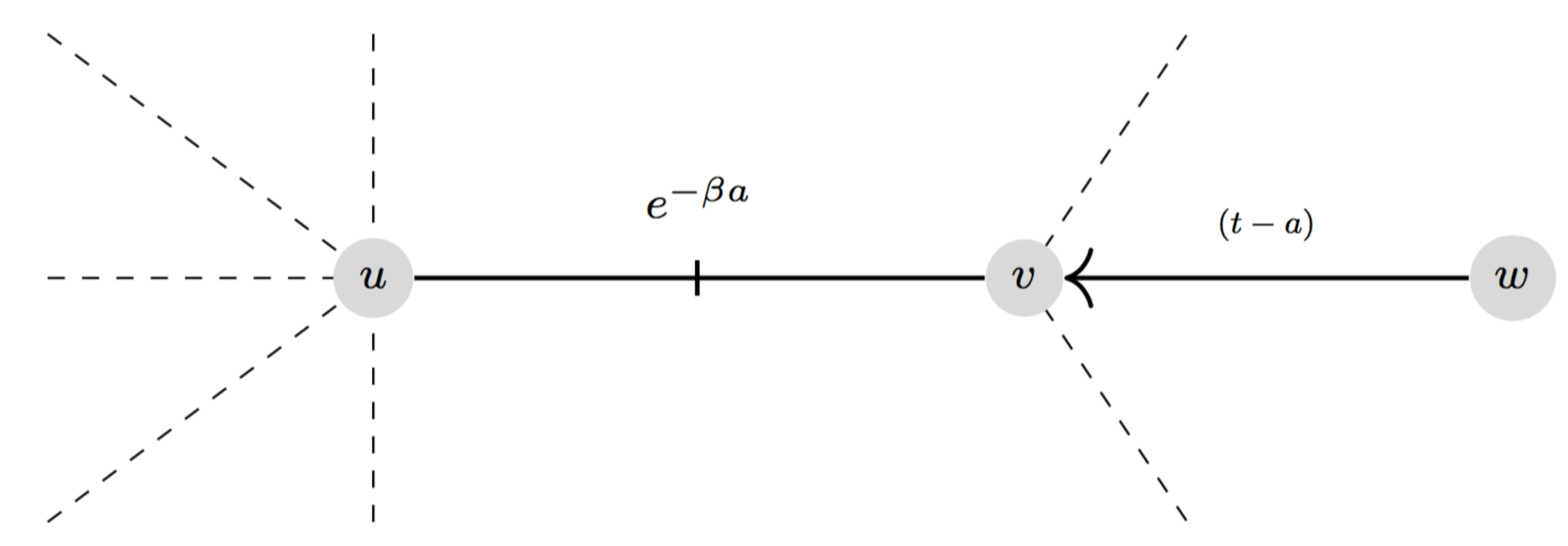}
	\caption{Consider the node labelled $u$ as the test node and thus in a cavity state. For its link with node $v$ to contribute to $\phi_I(t,a)$ it must be the case that $v$ received transmission from some neighbour $w$ at time $(t-a)$. If $t-a=0$ then this is equal to the initial proportion of infected nodes. Otherwise we take the probability of a transmission event $a$ time ago; which is $\beta \Phi_I(t-a)$. For $v$ to have been successfully infected at this time it must have been susceptible until that point, since two of its neighbours will not have transmitted before this time ($u$ is in a cavity state and $w$ will transmit at $(t-a)$) the probability of this is $zG_2(H_1(t-a))$ for $t > a$, illustrated by the dashed lines. Regardless of how neighbour $v$ was infected, the probability of $v$ not transmitting to $u$ before time $t$ is $e^{-\beta a}$ since the transmission process is Poisson. Finally, the neighbour $v$ must still be infected at age $a$, which is given by the survival function $\xi_q(a)$.}
	\label{fig:Tikz}
\end{figure}

To calculate $\diff{\{SI\}}{t}$ we first require $\diff{}{t} \Phi_I(t)$, from (\ref{eq:nonMarkEBCM}) we have
\begin{align*}
\diff{\Phi_I(t)}{t} &= \diff{}{t} \int_0^t \phi_I(t,a) \, da\\
&= \int_0^t \pd{}{t} \phi_I(t,a) \, da + \phi_I(t,t)\\
&= -\int_0^t \left[(\beta+\rho(a))\phi_I(t,a) + \pd{}{a} \phi_I(t,a) \right] \, da + \phi_I(t,t)\\
&= - \beta \Phi_I(t) + \phi_I(t,0) - \int_0^t \rho(a) \phi_I(t,a) \, da.
\end{align*}
Now to derive the evolution equation for $\{SI\}$ we start by differentiating both sides of~\eqref{eqn:curlySIdef}:
\begin{align*}
\diff{\{SI\}}{t} &= z N \ave{k} \diff{}{t} \left[G_1(\Theta(t)) \Phi_I(t)\right]\\
&= z N \ave{k} \left[ -\beta \Phi_I (t) G_2(\Theta(t)) \Phi_I (t) + G_1(\Theta(t)) \diff{}{t} \Phi_I(t)\right]\\
&= -\beta \frac{ \{SI\}\{SI\} G_2(\Theta)}{z N \ave{k}G_1(\Theta(t))G_1(\Theta(t))} + z N \ave{k} G_1(\Theta) \diff{}{t}\Phi_I(t)
\end{align*}
The first term already matches the first term of \eqref{eq:PLM:SI}.  We now explore the remaining terms
\begin{align*}
z N \ave{k} G_1(\Theta) \diff{}{t}\Phi_I(t) &=  z N \ave{k} G_1(\Theta(t))\left[- \beta \Phi_I(t) + \phi_I(t,0) - \int_0^t \rho(a) \phi_I(t,a) \, da \right],\\
&=-\beta \{SI\} + z N \ave{k} G_1(\Theta(t)) \left[ (1-z)\delta(t) \right.\\
&\hphantom{=-\beta S} \left. + z G_2(\Theta(t))\int_0^t \beta\phi_I(t,a) \, da - \int_0^t \rho(a) \phi_I(t,a) \, da  \right],\\
&= - \beta\{SI\} + z\beta \{SI\}G_2(\Theta(t)) \Phi_I(t) \\
&\hphantom{=-\beta SI+}- z N \ave{k} G_1(\Theta(t)) \int_0^t \rho(a) \phi_I(t,a) \, da.
\end{align*}
All that remains is to rewrite the final term in terms of the new variables. From~\eqref{eq:EBCM:phiDiff} we can use an integrating factor of $\exp\left(\int_0^a [\zeta(\hat{a}) + \rho(\hat{a})] \, da\right)$ to find
\begin{equation}
\phi_I(t,a) = e^{-\beta a} \xi_q(a) \phi_I(t-a,0),
\label{eq:IntFac}
\end{equation}
using the relation $\displaystyle{\xi_q(a) = \exp\left(-\int_0^a \rho(\hat{a}) \, da\right)}$ found in Lemma 1. We can progress further using~(\ref{eq:EBCM:phiI}) to give
\begin{align}
\phi_I(t,a) &= e^{-\beta a} \xi_q(a)   \left((1-z)\delta(t-a) +  zG_2(\Theta(t-a)) \int_0^{t-a} \beta \phi_I(t-a,\hat{a}) \, d\hat{a} \right) \nonumber\\
&= e^{-\beta a} \xi_q(a)  \left[(1-z)\delta(t-a) +  zG_2(\Theta(t-a)) \beta\Phi_I(t-a)\right]. \label{eq:phiIderive}
\end{align}
As an alternative, we offer a graphical description of (\ref{eq:phiIderive}) in Fig.~\ref{fig:Tikz}. Recall $q(a) = \rho(a) \xi_q(a)$, we have
\begin{align*}
\int_0^t \rho(a) \phi_I(t,a) \, da   &= \int_0^t \left\{\vphantom{\int} e^{-\beta a}q(a) \left[(1-z)\delta(t-a) \right. \right.\\
& \hphantom{+ze^{aa}} \left. \vphantom{\int}\left. + z\beta G_2(\Theta(t-a)) \beta \Phi_I(t-a)\right] \right\}\, da\\
&= \vphantom{\int} (1-z)e^{-\beta t} q(t) \\
& \hphantom{+z} + z\beta\int_0^t e^{-\beta a} q(a) G_2(\Theta(t-a))  \Phi_I(t-a)da.
\end{align*}
Substituting in $\Phi_I(t-a) = \{SI\}(t-a)/zN\ave{k}G_1(\Theta(t-a))$ gives
\begin{align*}
z N \ave{k} G_1(\Theta(t)) \int_0^t \rho(a) \phi_I(t,a) \, da   &= z (1-z)N \ave{k} G_1(\Theta(t))e^{-\beta t} q(t)\\
&\hspace{-2cm} + z G_1(\Theta(t)) \beta \int_0^tq(a) e^{-\beta a}\frac{G_2(\Theta(t-a))\{SI\}(t-a)}{G_1(\Theta(t-a))}\, da.
\end{align*}
Combining these results, the equation for $\diff{}{t} \{SI\}$ matches the equation for $\diff{}{t} [SI]$ if we replace $[SI]$ by $\{SI\}$ and $H_1$ by $\Theta$.

We now turn to the equation for $\{I\}$.  We will use
\begin{align*}
i(t,a) &= i(t-a,0)\xi_q(a)\\
& = (1-z)\delta(t-a)\xi_q(a) + z \ave{k}G_1(\Theta(t-a)) \xi_q(a)\int_0^{t-a} \beta \phi_I(t-a,\hat{a}) d\hat{a}\\
&= (1-z) \delta(t-a)\xi_q(a) + z \ave{k} G_1(\Theta(t-a)) \beta \Phi_I(t-a)\xi_q(a).
\end{align*}
\begin{align*}
\{I\} &= N\int_0^t i(t,a) \, da\\
&=N \int_0^t (1-z) \delta(t-a)\xi_q(a) + z \ave{k} G_1(\Theta(t-a)) \beta \Phi_I(t-a)\xi_q(a) \, da\\
&= N (1-z) \xi_q(t) + z N\ave{k} \beta  \int_0^t G_1(\Theta(t-a)) \Phi_I(t-a) \xi_q(a)\, da\\
&= N(1-z) \xi_q(t)  + \beta \int_0^t \{SI\}(t-a) \xi_q(a) \, da.
\end{align*}
Therefore, the expressions for $\{I\}$ and $[I]$ also coincide. Thus we have shown that the system of equations for $\{SI\}$, $\{I\}$, and $\Theta$ is identical to the system for $[SI]$, $[I]$, and $H_1$. This completes the proof.\hfill$\blacksquare$\\

\begin{figure}
\includegraphics[width = \textwidth]{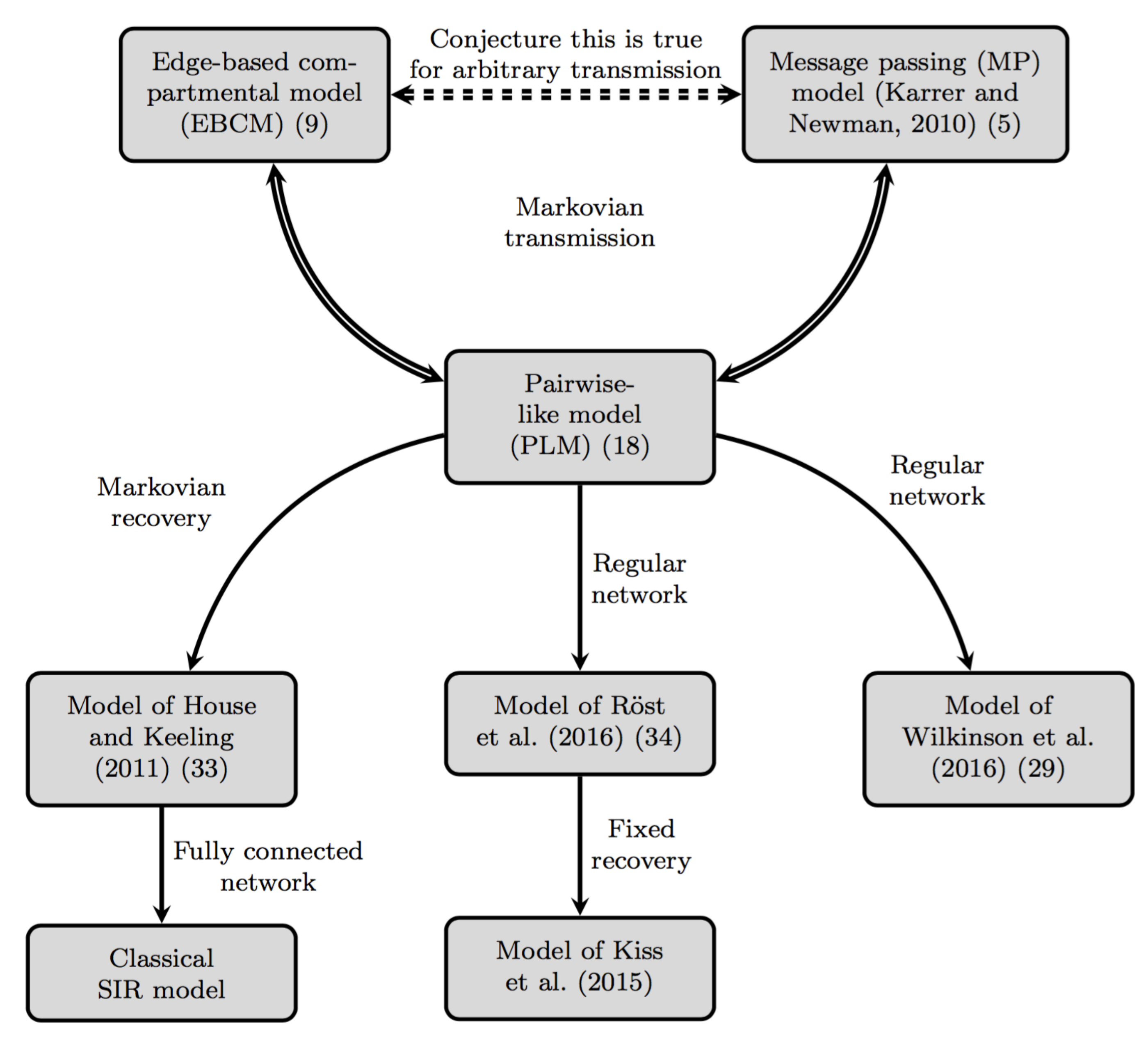}
	\caption{Diagram showing the relationship between the various models discussed in the paper. Under the assumption of Markovian/Poisson transmission, the PLM can be derived from the MP model as shown in Section \ref{sec:derive}, and the PLM and EBCM are equivalent by Theorem 2. From these proven equivalences, and the equivalence of $\Theta(t)$ and $H_1(t)$ under the most general circumstances, we conjecture that the MP and EBCM are equivalent for arbitrary transmission processes. Existing models can be derived as special cases of the PLM when the given additional assumptions are enforced, these are shown in Section~\ref{sec:regNet} and \ref{sec:recRed}.}
	\label{fig:mindmap}
\end{figure}

This equivalence shows that in the special case of Markovian transmission the EBCM (\ref{eq:nonMarkEBCM}) becomes exact on an ensemble of CM networks where the network size tends to infinity. Moreover, it suggests that the EBCM and MP models may indeed be equivalent for a general transmission process. The main step required to prove the general result is to show that $H_1$ and $\Theta$ satisfy the same evolution equation. To do this we take
\begin{align}
\diff{H_1}{t} &= - f(t) [1-z G_1(H_1(0))] - \int_0^t f(a) \left[ -z G_2(H_1(t-a))\diff{H_1(t-a)}{t} \right] \, da \nonumber\\
&= -f(t) [1-z G_1(1)] + \int_0^t f(a) \left[z G_2(H_1(t-a))\diff{H_1(t-a)}{t} \right] \, da \nonumber\\
&= -f(t)(1-z) + \int_0^t f(a) \left[z G_2(H_1(t-a))\diff{H_1(t-a)}{t} \right] \, da. \label{eq:generalH1}
\end{align}
The dynamics of $\Theta$ is governed by the following equation
\[
\diff{\Theta}{t} = - \int_0^t \zeta(a) \phi_I(t,a) \, da.
\]
Expression for $\phi_I(t,a)$ can be obtained from the generalised form of (\ref{eq:IntFac})
\[
\phi_I(t,a) = \phi_I(t-a,0) e^{\displaystyle -\int_0^a [\zeta(\hat{a}) + \rho(\hat{a})] \, d\hat{a}}.
\]
Introducing $\hat{f}(a) := \zeta(a) e^{\displaystyle -\int_0^a [\zeta(\hat{a}) + \rho(\hat{a})] \, d\hat{a}}$ and using
\[
\phi_I(t-a,0) = (1-z) \delta(t-a) - z G_2(\Theta(t-a)) \diff{\Theta(t-a)}{t},
\]
we have
\begin{align}
\diff{\Theta}{t}  &= -\int_0^t \hat{f}(a) \left[(1-z) \delta(t-a) - z G_2(\Theta(t-a)) \diff{\Theta(t-a)}{t} \right] da \nonumber\\
&= - \hat{f}(a) (1-z) + z\int_0^t \hat{f}(a)G_2(\Theta(t-a)) \diff{\Theta(t-a)}{t} \, da. \label{eq:generalTheta}
\end{align}
Thus, $H_1(t)$ and $\Theta(t)$ have the same dynamics if one can show that $f(a)=\hat{f}(a)$. From the definition of $f(a)$ and using the result of Lemma 1, we obtain
\[
	f(a) = \zeta(a) \xi_{\tau}(a) \xi_q(a)
		= \zeta(a) e^{\displaystyle -\int_0^a [\zeta(\hat{a}) + \rho(\hat{a})] \, d\hat{a}} = \hat{f}(a).
\]
Since $H_1$ and $\Theta$ have the same initial condition, this implies that $H_1(t)$ and $\Theta(t)$ will exhibit identical dynamics for general transmission and recovery processes. The equivalence between the models is illustrated in the top half of Fig.~\ref{fig:mindmap}.

\section{Derivation of existing models from the PLM for regular networks and different recovery time distributions}
\label{sec:reduction}
In this section we illustrate how the new PLM (\ref{eq:newSystem}) can be reduced to various existing models for appropriate choices of the recovery time and degree distributions. 
By illustrating this we emphasise that while the MP model and EBCM are probably the most competent models, the PLM is useful and plays the role of linking the most advanced
to the simpler models.
\subsection{Degree-regular networks}
\label{sec:regNet}
For a $k$-regular (homogeneous) network all nodes have the same degree, i.e. $k_u = \ave{k} = k$, and so the generating functions from (\ref{eq:genFunc}) simplify to 
\[ G_0(x) = x^k, \quad G_1(x) =  x^{k-1}, \quad \text{and} \quad G_2(x) = (k-1) x^{k-2}, \]
we also introduce two new variables
\begin{equation}
\begin{split}
	[S](t) &= zN G_0(H_1(t)) = zN[H_1(t)]^k, \\
	[SS](t) &= \ave{k} N \left(zG_1(H_1(t))\right)^2 =  kN\left(z[H_1(t)]^{k-1}\right)^2,
\end{split}
\label{eq:extraPairs}
\end{equation}
as the expected number of susceptible individuals, and the expected number of edges connecting two susceptible nodes, respectively. $[S](t)$ follows directly from (\ref{eq:msg_system}), $[SS](t)$ is defined by the number of edges connecting two nodes who were both initially susceptible at time $t=0$ and have escaped infection from their $(k-1)$ other neighbours.

Now we return to the PLM (\ref{eq:newSystem}) and the differential equation for $[I]$~(\ref{eq:PLMIdot}). Substituting in the simpler generating functions yields
\begin{equation}
\begin{split}
	\dot{H_1} &= -\beta \frac{[SI]}{z k N H_1^{k-1}}, \\
	\dot{[I]} &= \beta [SI] - \beta \int_0^t q(a) [SI](t-a) da - q(t)N(1-z), \\
	\dot{[SI]} &= -\beta [SI][SI]\frac{(k-1)H_1^{k-2}}{zkN [H_1^{k-1}]^2}  - \beta[SI] \\
	&\hspace{-0.3cm} + z \beta [SI](k-1)H_1^{k-2} -  q(t)e^{-\beta t}(1-z)z H_1^{k-1} k N\\
	&\hspace{-0.3cm} - z \beta \int_0^t q(a)e^{-\beta a} [SI](t-a)(k-1)[H_1(t-a))]^{k-2} \frac{[H_1(t)]^{k-1}}{[H_1(t-a)]^{k-1}} da.
\end{split}
\label{eq:regSystem1}
\end{equation}
This can be simplified further using (\ref{eq:extraPairs}), firstly noting that
\begin{equation}
	\frac{[SS]}{[S]} = \frac{N k z^2 H_1^{2(k-1)}}{N z H_1^k} = k z H_1^{k-2},
	\label{eq:SS/S}
\end{equation}
and, using $\dot{H_1}$ we see that
\begin{equation}
	\diff{(H_1^{k-1})}{t} = -\beta(k-1)\frac{[SI]}{z k N H_1^{k-1}} H_1^{k-2} = -\beta\frac{(k-1)}{k}\frac{[SI]}{[S]} H_1^{k-1}.
\end{equation}
Solving for $H_1^{k-1}$ in the ODE above, using separation of variables, leads to
\begin{equation}
	H_1^{k-1}(t) = \exp\left( -\beta\int_0^t \frac{(k-1)}{k}\frac{[SI](a)}{[S](a)}\, da \right).
	\label{eq:H1_ODE}
\end{equation}

The result of this is that the system no longer requires the message $H_1$; one can calculate the time derivatives of $[S]$ and $[SS]$ from (\ref{eq:extraPairs}). Using the new relations (\ref{eq:SS/S}) and (\ref{eq:H1_ODE}), system (\ref{eq:regSystem1}) can be rewritten to give
\begin{equation}
\begin{split}
	\dot{[S]} &= - \beta[SI], \\
	\dot{[I]} &= \beta [SI] - \beta \int_0^t q(a) [SI](t-a) da - q(t)N(1-z), \\
	\dot{[SS]} &= -2\beta\frac{(k-1)}{k}\frac{[SS][SI]}{[S]}, \\
	\dot{[SI]} &=  - \beta\frac{(k-1)}{k}\frac{[SI][SI]}{[S]}  - \beta [SI] + \beta \frac{(k-1)}{k}\frac{[SS][SI]}{[S]}  \\
	&\phantom{-\beta}  -  k Nq(t)e^{-\beta t}(1-z)z \exp\left( -\beta\int_0^t \frac{(k-1)}{k}\frac{[SI](a)}{[S](a)}da \right) \\
	&\phantom{-\beta} - \beta \int_0^t q(a)e^{-\beta a} \frac{(k-1)}{k}\frac{[SS](t - a)[SI](t-a)}{[S](t- a)} \mathcal{F}(t)da,
\end{split}
\label{eq:regSystemFinal}
\end{equation}
where
\begin{equation}
	\mathcal{F}(t) = \exp\left( -\beta\int_{t - a}^t \frac{(k-1)}{k}\frac{[SI](u)}{[S](u)} du\right).
\end{equation}
This is identical to the system proposed by Wilkinson et al. \cite{wilkinson2016relationships}. Recently, R\"ost et al. \cite{rost2016pairwise} considered the same problem, an SIR epidemic with Poisson transmission and an arbitrary distribution of the recovery time on a regular network. By constructing an age-structured system of PDEs they were able to reach a very similar more compact model. We have, therefore, shown that the PLM extends recent work to allow for an extra level of freedom by allowing general degree distributions to be modelled in a more compact model than previous methods allowed for.

\subsection{Special distributions of the recovery time}
\label{sec:recRed}
As mentioned previously, a popular choice for the recovery time distribution is to assume that times are exponentially distributed, i.e. $q(a) = \gamma e^{-\gamma a}$ for $\gamma >0$, where $1/\gamma$ is the mean duration of infection. We briefly explain how this assumption simplifies the model and leads to familiar or well-known models. When this choice for $q(a)$ is substituted into (\ref{eq:PLMIdot}), we have
\begin{equation}
	\dot{[I]} = \beta[SI] - \gamma \left[\int_0^t e^{-\gamma a} \beta[SI](t-a) da + e^{-\gamma t}N(1-z)\right].
	\label{eq:markovI}
\end{equation}
Note that $e^{-\gamma a}$ is the probability of an infected node not recovering before age $a$, and since the number of infected nodes created $a$ time ago is $\beta[SI](t-a)$ for $a < t$ and $N(1-z)$ is the number of initially infected nodes then (\ref{eq:markovI}) can be rewritten as
\begin{equation}
	\dot{[I]} = \beta[SI] - \gamma [I].
\end{equation}
A similar result occurs when exponentially distributed recovery times are used in (\ref{eq:PLM:SI}), in which case the extra terms in the integral describe the probability for the susceptible node of an $[SI]$ edge to have survived until age $a$ without receiving transmission, either along this edge or another infected neighbour. Therefore, by the same logic one can replace the final two terms in (\ref{eq:PLM:SI}) with $\gamma [SI]$. This leads to a model, which, although formulated differently, is similar to models of Volz \cite{volz2008sir} and House and Keeling \cite{house2011insights}, namely, 
\begin{equation}
\begin{split}
	\dot{H_1} &= -\beta \frac{[SI]}{z \ave{k} N G_1(H_1)}, \\
	\dot{[I]} &= \beta [SI] - \gamma[I], \\
	\dot{[SI]} &= -\beta [SI][SI]\frac{G_2(H_1)}{z\ave{k}N [G_1(H_1)]^2}  - (\beta + \gamma)[SI] + z \beta [SI]G_2(H_1).
\end{split}
\label{eq:newSystemMarkov}
\end{equation}
Early pairwise models assumed that the contact network was regular, and that transmission and recovery times were exponentially distributed \cite{keeling1997disease}. Applying these assumptions to (\ref{eq:regSystemFinal}) leads to recovering such early models.

We examine another special case, when the duration of infection is a fixed period of time, $\sigma$, so that $q(a) = \delta(a - \sigma)$. When this is substituted into the PLM (\ref{eq:newSystem}), the integral terms are non-zero only at the point $a = \sigma$, since as soon as a node is infected at time $t_1$, it is known that this node will recover at exactly $t_2 = t_1 + \sigma$. This means that the system of integro-differential equations simplifies to a delay differential equation model, as stated below
\begin{equation}
\begin{split}
	\dot{H_1} &= -\beta \frac{[SI]}{z \ave{k} N G_1(H_1)}, \\
	\dot{[I]} &= \beta [SI] - \beta [SI](t-\sigma) - \delta(t-\sigma)N(1-z), \\
	\dot{[SI]} &= -\beta [SI][SI]\frac{G_2(H_1)}{z\ave{k}N [G_1(H_1)]^2}  - \beta[SI] \\
	&\phantom{-\beta} + z \beta [SI]G_2(H_1) -  \delta(t - \sigma)e^{-\beta t}(1-z)z G_1(H_1) \ave{k} N\\
	&\phantom{-\beta} -  z \beta e^{-\beta \sigma} [SI](t-\sigma)G_2(H_1(t-\sigma)) \frac{G_1(H_1(t))}{G_1(H_1(t-\sigma))}.
\end{split}
\label{eq:newSystemDelay}
\end{equation}
This model generalises the recent work of Kiss et al. \cite{kiss2015pairwise} to heterogeneous networks, and once again the original model in that paper can be retrieved when $q(a)$ is chosen to be a delta distribution in (\ref{eq:regSystemFinal}) (that original model did not explicitly account for the recovery of initially infected nodes). We illustrate all the model equivalences and reductions in Fig. \ref{fig:mindmap}.

Finally, it is worth briefly noting that in the case of a fully connected network, corresponding to a homogeneously well-mixed population we have that $[SI] = [S][I]$ and thus the earliest models, which assumed that the population was unstructured, can be recovered.

\section{Numerical simulation of the Pairwise-like Model}
\label{sec:sims}
In order to illustrate the accuracy of the newly derived PLM (\ref{eq:newSystem}) we compare the numerical solution of this model to results of direct stochastic network simulation. A common approach for simulating traditional Markovian models has been to use the Gillespie algorithm \cite{gillespie1977exact}, however, as modelling started to move away from the purely Markovian models, novel stochastic simulation methods have been derived \cite{anderson2007modified,boguna2014simulating} which provide efficient simulation algorithms that are able to generate true sample paths of the stochastic process. In this section we take advantage of the fact that transmission remains a Poisson process in order to use an algorithm similar to those described by Barrio et al. \cite{barrio2006oscillatory}. This approach is sometimes known as the rejection method and was proven to be stochastically exact by Anderson \cite{anderson2007modified}. The transmission process is run as in the standard Gillespie algorithm, and when a node becomes infected, a recovery time is drawn from the distribution $q(a)$; at each time step the time of next transmission is randomly calculated. However, if an infected node is scheduled to recover sooner then the next planned transmission event is rejected, and the new time is updated to the next recovery time (for full details see \cite{anderson2007modified}).

\begin{figure}
	\includegraphics[width = \textwidth]{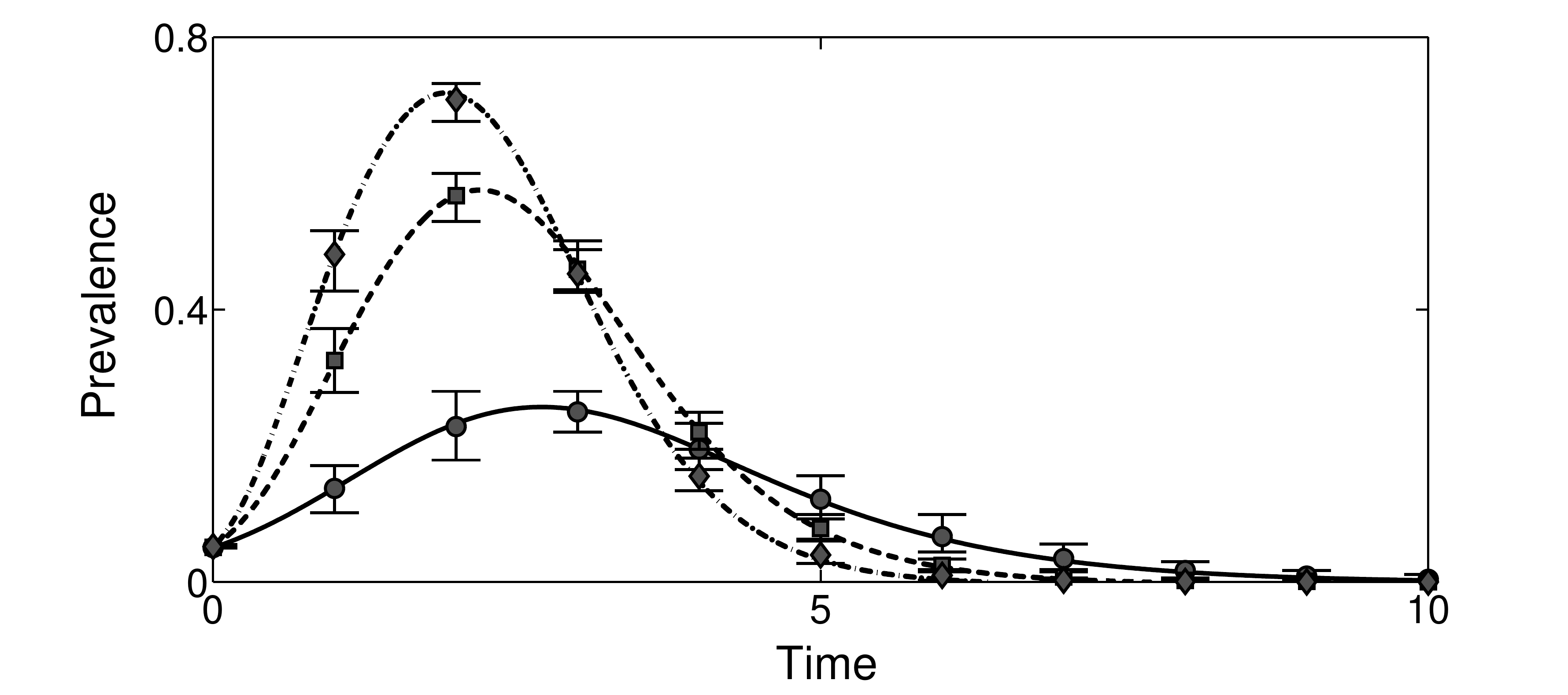}
	\caption{A plot comparing the output from the PLM (\ref{eq:newSystem}) for an epidemic where the recovery time distribution is normal with mean $2$ and standard deviation $0.5$. 
	The transmission parameter takes values $\beta = 0.15$ (solid line, circles), $0.3$ (dashed line, squares) and $0.4$ (dot-dashed line, diamonds),
	the markers represent the average of 100 simulations with corresponding parameters. The underlying networks of 1000 nodes are truncated scale-free networks with exponent $2.5$ and degree bounded between $3$ and $60$. Error bars denote the 5th and 95th percentile.}
	\label{fig:simsPLM}
\end{figure}

\begin{figure}
	\includegraphics[width = \textwidth]{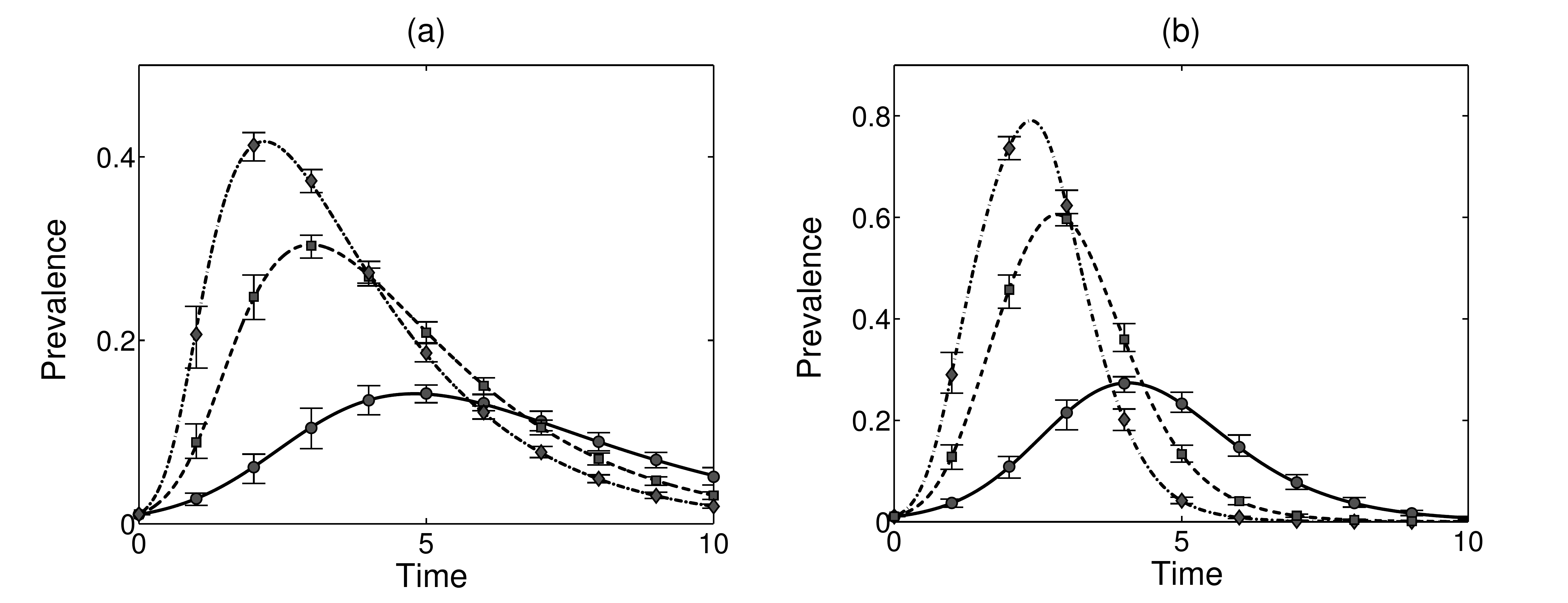}
	\caption{Tests of the models (\ref{eq:newSystemMarkov}) and (\ref{eq:newSystemDelay}) on truncated scale-free networks of 10000 nodes with exponent $2.5$ and degree bounded between $3$ and $60$. The transmission parameter takes values $\beta = 0.15$ (solid line, circles), $0.3$ (dashed line, squares) and $0.4$ (dot-dashed line, diamonds). In (a) results from (\ref{eq:newSystemMarkov}) are compared to numerical simulations, with $\gamma=0.5$ for all results. In (b) (\ref{eq:newSystemDelay}) is tested against numerical simulations with $\sigma = 2$. Results based on simulations are given by markers. Error bars denote the 5th and 95th percentile.}
	\label{fig:simsreduction}
\end{figure}

In the very early stages of an outbreak stochastic effects dominate the dynamics of the epidemic spread, which means that numerical simulations can often produce results that differ greatly from deterministic predictions. In this region methods such as branching process approximations \cite{heesterbeek2000mathematical} are more appropriate. To ensure that this does not affect our results we allow every iteration of the algorithm to reach a point where the stochastic effects are no longer a concern, and the infected population behaves deterministically. In practice this is achieved by running each individual realisation of the epidemic from a single initial seed until a specified level of infectivity is reached, at which point time is reset to zero in both the simulation and the PLM. A sufficient number of individual simulations are averaged to ensure that the mean behaviour of the stochastic model is correctly captured and is suitable for comparison with results derived from the deterministic or mean-field models. The results of such comparisons are shown for different transmission rates and various choices of the recovery time distribution in Figs.~\ref{fig:simsPLM} (normally distributed infection times) and~\ref{fig:simsreduction} (exponentially distributed and fixed recovery times). Firstly, in Fig.~\ref{fig:simsPLM} the full model (\ref{eq:newSystem}) is compared to numerical simulation on scale-free networks with normally distributed recovery times. The exceptional agreement between the simulation and the mean-field model in all cases shows strong visual evidence of this model's merits. In Fig.~\ref{fig:simsreduction} we repeat these tests for models (\ref{eq:newSystemMarkov}) and (\ref{eq:newSystemDelay}), which again show excellent agreement with numerical simulations. 
The agreement for networks with high degree heterogeneity and infection time distribution ranging from exponential and normal to fixed highlight the flexibility of this model.

The transmission parameters and the mean duration of infection are identical across all figures, however, the dynamics of the epidemics vary greatly in response to the change in recovery time distribution. In Fig.~\ref{fig:simsreduction} it is clear that a fixed duration of infection causes a much larger outbreak when compared to the case of exponentially distributed recovery times, despite both of them having the same mean duration of infection. This is largely due to the change in the variance of the recovery times. In Fig.~\ref{fig:simsreduction} (a) the exponentially distributed recovery time has the variance equal to $\gamma^{-2} = 4$, whereas, in Fig.~\ref{fig:simsreduction} (b) the variance in recovery time is zero. This highlights the need for the true characteristics of the diseases to be accurately studied and modelled if accurate and reliable predictions are to be achieved.

\section{Discussion}
\label{sec:discuss}

In this paper we have reviewed the message passing formalism for SIR epidemics on networks, and introduced a novel extension of the edge-based compartmental model to the case of arbitrary transmission and recovery processes. Both of these models are theoretically capable of accurately describing the expected dynamics of non-Markovian epidemics on tree networks; although in some cases it may be challenging to find a numerically solvable expression for the message $H_1$. The final epidemic size formula found using the EBCM is in perfect agreement with results reported in existing literature \cite{kenah2007second,karrer2010message}.

Adapting recent methods \cite{wilkinson2014message,wilkinson2016relationships} enabled us to construct a pairwise-like model (\ref{eq:newSystem}) when the transmission process is Markovian. This compact model of only two differential equations is easier to implement numerically compared to the new EBCM and still provides an exact description of an SIR epidemic with general recovery time and degree distribution for CM networks in the limit as $N \to \infty$. Therefore, by proving that the PLM is equivalent to the EBCM we have also shown that the EBCM is equivalent to the MP model when the transmission is Markovian. We believe that this equivalence holds in a general setting, and have proved it for the message quantities, $H_1$ and $\Theta$.

This new PLM model is more flexible than other pairwise models to date and thus can be applied to a wider range of data gathered from field studies. Unlike many other pairwise models, the PLM has been derived rather than heuristically defined, as is often the case \cite{eames2002modeling,gross2006epidemic,house2011insights}. Furthermore, the presence of terms typically found when using the message passing formalism means that our PLM circumvents the issue of dependence on higher order arrangements, and the closure of these as found in most pairwise-like models. As a result, epidemics on networks which have a highly heterogeneous degree distribution can often lead to large systems which require many differential equations to be solved \cite{eames2002modeling}, or complex moment closures \cite{simon2015super}. However, obtaining estimates on the accuracy of such approximations remains a challenge \cite{pellis2015exact}. Essentially, the new PLM is a hybrid of classical pairwise and MP models.

The PLM has been studied under specific choices for both the node degree and recovery time distribution, and it has been shown that many recent and classical models can be extracted from it. By demonstrating this we hope to provide some intuition for how the PLM works and to illustrate that these newest models build on existing models but provide a modern twist. It is encouraging that such mean-field models remain relatively compact highlighting that the SIR epidemic can be modelled quite effectively, as long a small number of key indicators about the network and the epidemic process are known. Our theoretical results have been supported by an excellent agreement between mean-field and the exact stochastic models showing that these models can be practically implemented and numerically solved, giving an excellent characterisation of the average stochastic behaviour.

Numerous extensions of the present work are possible. For example, the implementation of an efficient solver of the novel EBCM is still outstanding. Efficient numerical methods to solve such age-structured models exist but this was outside the scope of our study. In some sense the novel EBCM is the most complete mean-field model when one considers SIR epidemics on CM networks. This is due to the new model being able to handle arbitrary degree distributions, as well as arbitrary transmission and recovery processes. Additionally, such models could be refined to account for dynamic or adaptive contacts. Dynamic networks have already been incorporated in edge-based modelling in the purely Markovian setting \cite{miller2012edge}, and it may be possible to extend this to a more general framework to allow for a much more unified treatment of models that include the concurrent spread of the diseases and link turnover.

\section*{Acknowledgements}
N. Sherborne acknowledges funding for his PhD studies from the EPSRC (Engineering and Physical Sciences Research Council), EP/M506667/1, and the University of Sussex.

\bibliographystyle{unsrt}
\bibliography{3bibv2}

\end{document}